# High-pressure synthesis of MnO–ZnO solid solutions with rock salt structure: *in situ* X-ray diffraction studies

Petr S. Sokolov,[1,2] Andrey N. Baranov,[2] Christian Lathe,[3] Vladimir Z. Turkevich,[4] Vladimir L. Solozhenko [1*]

[1] *LPMTM–CNRS, Université Paris Nord, 93430 Villetaneuse, France*
[2] *Chemistry Department, Moscow State University, 119991 Moscow, Russia*
[3] *HASYLAB–DESY, 22603 Hamburg, Germany*
[4] *Institute for Superhard Materials, 04074 Kiev, Ukraine*

**Abstract**

X-ray diffraction with synchrotron radiation has been used for the first time to study chemical interaction in the MnO–ZnO system at 4.8 GPa and temperatures up to 1600 K. Above 750 K, the chemical reaction between MnO and ZnO has been observed that resulted in the formation of rock salt (*rs*) $Mn_{1-x}Zn_xO$ solid solutions ($0.3 \leq x \leq 0.7$). The lattice parameters of these solid solutions have been *in situ* measured at high pressure as a function of temperature, and corresponding thermal expansion coefficients have been calculated.

* e-mail: vladimir.solozhenko@univ-paris13.fr

**Introduction**

Zinc oxide is a wide-band-gap semiconductor and has many industrial applications [1,2]. Recently ZnO-based solid solutions with transition metal oxides have gained substantial interest as promising magnetic semiconductors [2]. At ambient pressure ZnO has hexagonal wurtzite structure ($P6_3mc$) that transforms into rock salt one ($Fm3m$) at pressures above 6 GPa [3,4]. High-pressure phase of ZnO is kinetically stable only above 2 GPa and cannot be quenched down to ambient conditions [4]. Very recently, metastable $Me^{II}O$-ZnO solid solutions ($Me^{II}$ – $Ni^{2+}$, $Fe^{2+}$, $Co^{2+}$, $Mn^{2+}$) with rock salt structure have been synthesized by quenching from 7.7 GPa and 1450-1650 K [5]; in particular, single-phase rock salt $Mn_{1-x}Zn_xO$ (MnZO) solid solutions of various stoichiometries ($0 < x \leq 0.4$) have been obtained. The thermodynamic study of the MnO-ZnO system have shown that at ambient pressure there is a wide (from 10 to 75 mol% ZnO at 1073 K) region of coexistence of rock salt MnO-based and wurtzite ZnO-based solid solutions [6]. However, no information on chemical interaction and phase relations in the system at high pressures and temperatures can be found in the literature. In the present work we have performed the first *in situ* study of the binary MnO-ZnO system at high pressure and temperatures using X-ray diffraction with synchrotron radiation.

**Experimental**

High-purity powders of wurtzite ZnO (Alfa Aesar, 99.99%, 325 mesh) and rock salt MnO (Alfa Aesar, 99.99%, 325 mesh) have been used as starting materials. The MnO-ZnO mixtures of various compositions (0, 30, 50 and 70 mol% ZnO) were thoroughly ground in a mortar, compressed into disks and placed into capsules of high-purity hexagonal graphite-like boron nitride (hBN).

High-pressure experiments have been performed using MAX80 multianvil X-ray system at beamline F2.1, DORIS III (HASYLAB-DESY). The experimental details and high-pressure setup have been described elsewhere [7,8]. Energy-dispersive X-ray diffraction patterns were collected on a Canberra solid state Ge-detector with fixed Bragg angle $2\theta = 9.073(1)°$ using a white beam collimated down to $100 \times 100$ μm$^2$. The detector was calibrated using the $K_\alpha$ and $K_\beta$ fluorescence lines of Rb, Mo, Ag, Ba, and Tb.

The sample temperature was measured by a Pt10%Rh–Pt thermocouple. The correction for the pressure effect on the thermocouple emf was made using the data of Getting and Kennedy [9] extrapolated to 5 GPa. The temperature in the high-pressure cell was controlled by a Eurotherm PID regulator within ±3 K. Pressure at various temperatures were evaluated from the lattice parameters of highly ordered ($P_3 = 0.98 \pm 0.02$)[1] hBN using its thermoelastic equation of state [7]. With the temperature growth up to 1600 K, the pressure increase in the central part of the cell did not exceed 0.2 GPa.

---

[1] ($P_3 = 1-\gamma$), where $\gamma$ is the concentration of turbostratic stacking faults [10]

The samples were compressed to the required pressure at ambient temperature, and then diffraction patterns were collected in the "autosequence" mode at a linear heating with a rate of 10 K/min. With the storage ring operating at 4.44 GeV and 120±30 mA, the time of data collection for each pattern was 60 s.

**Results and discussion**

At 4.8 GPa and temperatures below 730 K only reflections of pristine oxides (*w*-ZnO and *rs*-MnO) are observed in the diffraction patterns. Appearance of new reflections corresponding to the *rs*-Mn$_{1-x}$Zn$_x$O solid solutions has been observed at temperatures starting from ~735 K (see Table 1). Regardless of the stoichiometry, the lattice parameters of the forming solid solutions are very different from the lattice parameter of pristine *rs*-MnO (Table 2), contrary to the FeO-ZnO system [11]. Upon further temperature increase, intensities of *w*-ZnO and *rs*-MnO reflections start to decrease indicating chemical interaction in the system. This reaction is accompanied by dissolution of zinc oxide in manganese (II) oxide and formation of a Mn$_{1-x}$Zn$_x$O solid solution with rock salt structure. The forming MnZO solid solution coexists with *w*-ZnO and *rs*-MnO in a rather wide temperature range. The temperature of compete disappearance of *w*-ZnO ($T_d$) depends on the initial stoichiometry as shown in Table 1 and increases with ZnO content from 943±5 K for $x = 0.3$ to 1243±5 K for $x = 0.7$. At higher temperatures ($>T_d$) only reflections of *rs*-Mn$_{1-x}$Zn$_x$O (as main phase) and *rs*-MnO (as impurity) are observed in the diffraction patterns. The temperature of complete disappearance of *rs*-MnO ($T_{cd}$) varies from 1100 to 1350 K depending of the stoichiometry (see Table 1), and finally only reflections of rock salt MnZO phase remain in the diffraction patterns (Fig. 1). The compositions of the as-formed *rs*-MnZO solid solutions thus attain the compositions of the initial reaction mixtures.

Quenching down to ambient conditions results in formation of single-phase *rs*-Mn$_{1-x}$Zn$_x$O solid solutions only at relatively low ZnO concentration i.e. $x \leq 0.4$. The recovered samples with higher ZnO content are two-phase mixtures of MnZO solid solutions with rock salt and wurtzite structures which is in a good agreement with our previous results on synthesis of MnO–ZnO solid solutions by quenching from 7.7 GPa [5].

Likewise to the FeO–ZnO system [11], at temperatures above 1400 K, drastic changes in intensities of reflections are observed in diffraction patterns for all *rs*-MnZO solid solutions.[2] Such fluctuations of intensities can be caused by the inferred motion of crystallites due to appearance of a liquid in the system that is typical for an onset of melting. It should be noted, that at ambient pressure melting points of *w*-ZnO and *rs*-MnO are 2248±25 K [12] and 2123±20 K [13], respectively, but $dp/dT$ slopes for their melting curves are not known. In the case of *rs*-Mn$_{0.5}$Zn$_{0.5}$O, at 4.7±0.2 ГПа we observed the complete melting at 1643±20 K.

---

[2] For *rs*-Mn$_{0.7}$Zn$_{0.3}$O an abrupt (more than by one order of magnitude) increase in intensity of *220* reflection is observed, while *331* and *400* reflections disappear almost completely; for Mn$_{0.3}$Zn$_{0.7}$O, *311* and *420* reflection intensities increase twice, and intensity of *220* and *511* reflection decreases. No changes in reflection intensities are observed for pristine manganese (II) oxide at the same *p,T*-conditions.

Lattice parameters of $rs$-Mn$_{1-x}$Zn$_x$O solid solutions ($x$ = 0.3, 0.5, 0.7) have been determined in the T$_d$ – 1500 K range at 4.8 GPa. The linear temperature dependencies of the unit cell volumes are shown in Fig. 2, and their slopes give the values of the volume thermal expansion coefficients ($\alpha$) at 4.8 GPa (Table 2). The thermal expansion coefficient for pure manganese (II) oxide, 3.3(4)×10$^{-5}$ K$^{-1}$, is in a good agreement with the value 3.45(1)×10$^{-5}$ K$^{-1}$ reported in [14]. Linear extrapolation of the thermal expansion coefficients of $rs$-Mn$_{1-x}$Zn$_x$O solid solutions to $x$ = 1 ($rs$-ZnO) gives the value of 8.2(9)×10$^{-5}$ K$^{-1}$.

**Conclusions**

At pressures of about 5 GPa and temperatures above 750 K chemical interaction between MnO and ZnO results in the formation of Mn$_{1-x}$Zn$_x$O (0.3 ≤ $x$ ≤ 0.7) solid solutions with rock salt structure. Temperature of formation of single-phase $rs$-Mn$_{1-x}$Zn$_x$O increases with increasing ZnO content from 943 K for $x$ = 0.3 to 1243 K for $x$ = 0.7. Thermal expansion coefficients calculated from the temperature dependencies of lattice parameters of $rs$-Mn$_{1-x}$Zn$_x$O solid solutions under pressure are intermediate between corresponding values of pristine MnO and ZnO.

**Acknowledgements**

Experiments have been performed during beamtime allocated to the Project I-20090172 EC at HASYLAB-DESY and have received funding from the European Community's Seventh Framework Programme (FP7/2007-2013) under grant agreement n° 226716. This work was also supported by Russian Foundation for Basic Research (Project No 09-03-90442-Укр_ф_а) and Ukrainian Foundation for Basic Research (Project No Ф28.7/008). PSS is grateful to the French Ministry of Foreign Affairs for financial support (BGF fellowship No 2007 1572).

**Table 1.**

Characteristic temperatures of chemical interaction in the MnO–ZnO system at 4.8 GPa

| Temperature* | ZnO/(MnO+ZnO) molar ratio | | |
|---|---|---|---|
| | **0.3** | **0.5** | **0.7** |
| $T_s$ (K) | 733 | 743 | 803 |
| $T_d$ (K) | 943 | 1043 | 1243 |
| $T_{cd}$ (K) | 1093 | 1243 | 1343 |

\* $T_c$ is onset temperature of *rs*-MnZO formation; $T_d$ is temperature of complete disappearance of *w*-ZnO, and $T_{cd}$ is temperature of complete disappearance of *rs*-MnO; in all cases the error is ±5 K

**Table 2.**

Lattice parameters and volume thermal expansion coefficients of rock salt MnO-ZnO solid solutions at 4.8 GPa

| **Composition** | **MnO** | **$Mn_{0.7}Zn_{0.3}O$** | **$Mn_{0.5}Zn_{0.5}O$** | **$Mn_{0.3}Zn_{0.7}O$** | ***rs*-ZnO\*** |
|---|---|---|---|---|---|
| **Lattice parameters at 1298 K (Å)** | 4.450(4) | 4.415(3) | 4.376(4) | 4.349(5) | 4.305 |
| **α×10$^5$ (K$^{-1}$)** | 3.36±0.37 | 6.96±0.33 | 7.4±0.45 | 7.69±1 | 8.2±0.9 |
| **Temperature range (K)** | 800-1500 | 943-1500 | 1043-1500 | 1243-1500 | – |

\* Linear extrapolation of the thermal expansion coefficients and lattice parameters of *rs*-$Mn_{1-x}Zn_xO$ (*x* = 0, 0.3, 0.5, 0.7) solid solutions to *x* = 1 (*rs*-ZnO)

**Figure captions**

Fig. 1.  Diffraction patterns of the MnO-ZnO mixture (50 mol% ZnO) taken at 4.8 GPa in the course of linear heating at a rate of 10 K/min.

Fig. 2.  Unit cell volumes of rock salt $Mn_{1-x}Zn_xO$ solid solutions *vs* temperature at 4.8 GPa (squares – $x = 0$; circles – $x = 0.3$; triangles – $x = 0.5$; hexagons – $x = 0.7$).

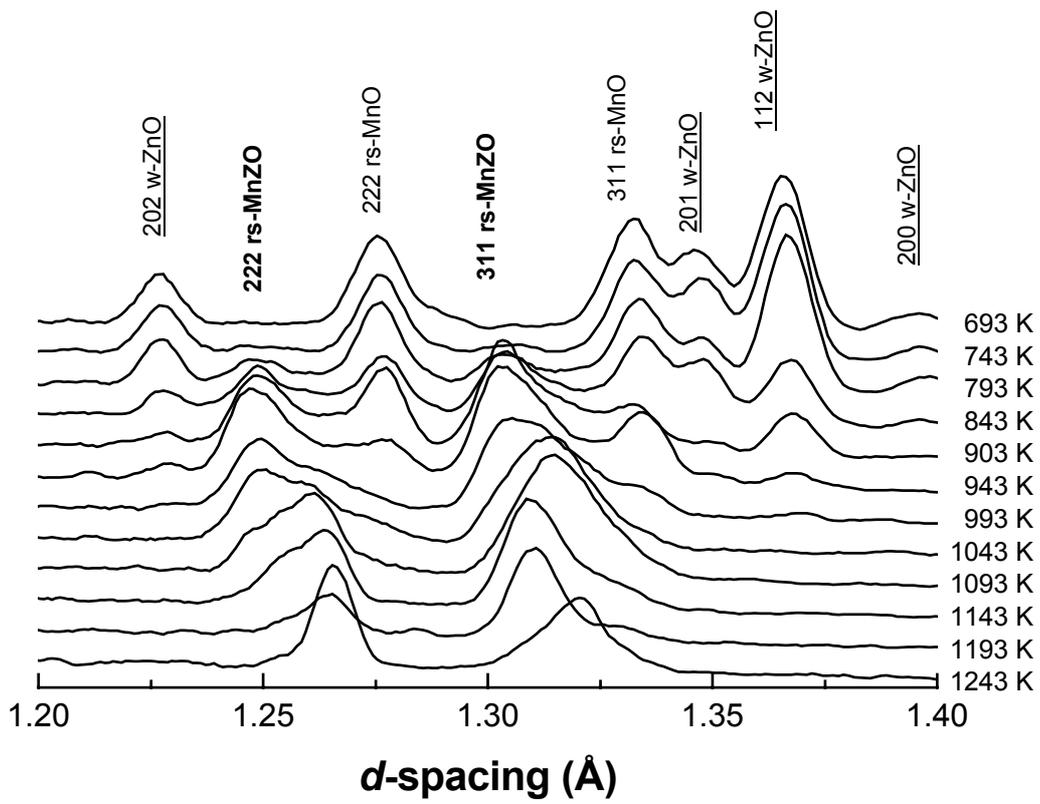

**Fig. 1**

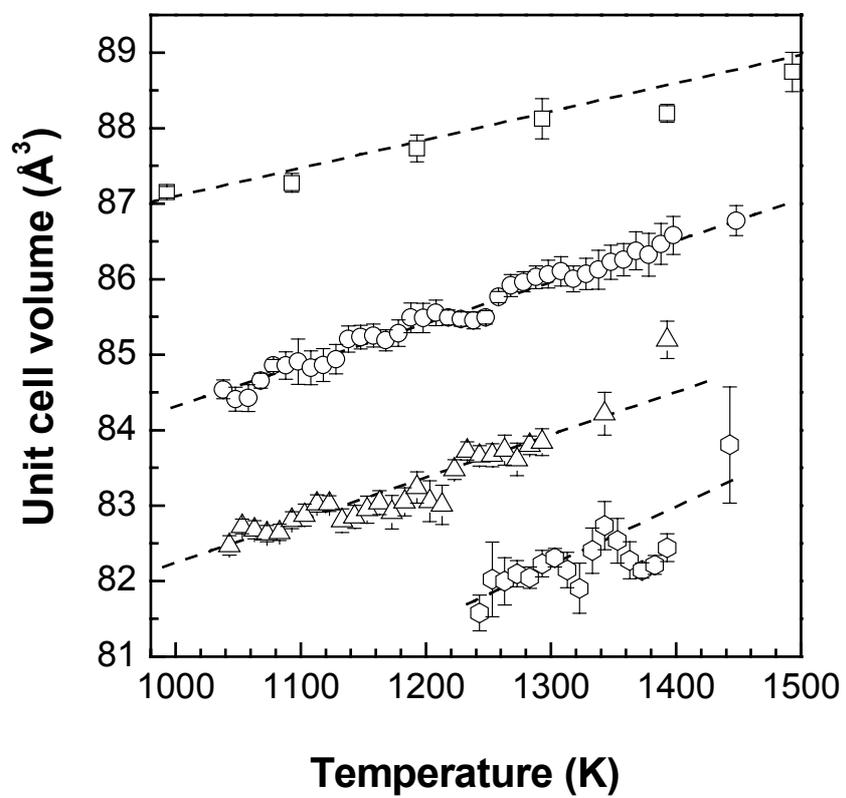

Fig. 2